\documentclass[aps,pra,twocolumn,superscriptaddress,eps2pdf,notitlepage]{revtex4-2}

 \usepackage{graphicx} 
\usepackage{booktabs}
 \usepackage{epsfig}
 \usepackage{amsfonts}
\usepackage{amssymb}
 \usepackage{mathrsfs}
\usepackage{theorem}
\usepackage{amsmath}
 \usepackage{soul}
 \usepackage{cleveref}

\usepackage{color,dsfont}
\newtheorem{theorem}{Theorem}

\newtheorem{lem}{Lemma}
\newtheorem{corollary}{Corollary}

\newcommand{\ket}[1]{\ensuremath{\vert#1\rangle}}

\newcommand{\etc}[1]{#1}  

\def\id{\mbox{\small 1} \!\! \mbox{1}}

\begin{document}
\title{Early fault-tolerant simulations of the Hubbard model}

\author{Earl T.\ Campbell}
\affiliation{AWS Center for Quantum Computing, Pasadena, CA 91125 USA}

\begin{abstract}
Simulation of the Hubbard model is a leading candidate for the first useful applications of a fault-tolerant quantum computer. A recent study of quantum algorithms for early simulations of the Hubbard model [Kivlichan \textit{et al.} Quantum 4 296 (2019)] found that the lowest resource costs were achieved by split-operator Trotterization combined with the fast-fermionic Fourier transform (FFFT) on an $L \times L$ lattice with length $L=2^k$.  On lattices with length $L \neq 2^k$, Givens rotations can be used instead of the FFFT but lead to considerably higher resource costs. We present a new analytic approach to bounding the simulation error due to Trotterization that provides much tighter bounds for the split-operator FFFT method,  leading to $16 \times$ improvement in error bounds. Furthermore, we introduce plaquette Trotterization that works on any size lattice and apply our improved error bound analysis to show competitive resource costs.  We consider a phase estimation task and show plaquette Trotterization reduces the number of non-Clifford gates by a factor $5.5\times$ to $9 \times$ (depending on the parameter regime) over the best previous estimates for $8 \times 8$ and $16 \times 16$ lattices and a much larger factor for other lattice sizes not of the form $L=2^k$.  In conclusion, we find there is a potentially useful application for fault-tolerant quantum computers using around one million Toffoli gates.
\end{abstract}

\maketitle  

Free, non-interacting, fermionic systems can be efficiently solved on a classical computer and occasionally by pen and paper calculation, thereby determining their exact energy spectrum.  In contrast, even simple interacting electronic systems can prove intractable.  Hubbard proposed a simple model of interacting electrons that describes the physics of real world systems of interest~\cite{hubbard1963electron} and might elucidate the mechanisms behind high-temperature superconductivity. Nevertheless, it is sufficiently complex that we can not currently simulate the Hubbard model for large lattices at high accuracy.  An important quantity is the ground state energy per lattice site in the two dimensional square lattice (TDL).  Density matrix renormalization group approximates the TDL with finite cylinders
of widths of up to 7 sites and lengths of up to 64 sites~\cite{zheng2017stripe}.  This technique has an uncontrolled, systematic error due to finite width.  However, using a variety of simulation techniques~\cite{zheng2017stripe}, each with their own source of uncontrolled error, there is agreement up to $\pm 0.5 \%$ in the ground state energy density in the weak doping regime.  This error is regarded as the aggregated accuracy of current classical methods and achieving better accuracy is extremely challenging. From a complexity theory perspective, the ability to find the ground states of any Hubbard model Hamiltonian~\cite{childs2014bose} would enable a solution of any problem in the complexity class Quantum-Merlin-Arthur (QMA).

\etc{The Hubbard model is one of the earliest examples where quantum algorithms have been proposed~\cite{AbramsHubbard}. Recently,} the Hubbard model has been identified as a potential application for noisy, near term quantum computers~\cite{cade2019strategies,clinton2020hamiltonian,bauer2020quantum}.  However, these are heuristic algorithms without performance guarantees, and noise may prohibit them from achieving the required accuracy to be competitive with classical simulations.  Building a fault-tolerant device is a longer term goal, but offers rigorous performance guarantees and gentle resource scaling with respect to accuracy. Kivlichan \textit{et al}~\cite{kivlichan2020improved} considered estimation of the ground state energy of the Hubbard TDL using \etc{phase estimation~\cite{Kitaev97,PhaseEstimationOld}} and second-order Trotter formulae.  For instance, for an $8 \times 8$ Hubbard simulation at $0.5 \%$ relative error, they estimated that a transmon-based surface code architecture may require only 62,000 physical qubits and 23 million non-Clifford $T$ gates.  While such a device would be much larger than any currently existing, these resource estimates are orders of magnitude smaller than those needed to break RSA~\cite{Gorman2017quantum,gidney2019factor}, or even more costly, to perform combinatorial optimisation~\cite{campbell2019applying,sanders2020compilation}.

For this task, Kivlichan \textit{et al}~\cite{kivlichan2020improved} considered several variants of second-order Trotter~\cite{suzuki1990fractal,berry2007efficient,babbush2015chemical,childs2018toward,childs2019theory} to approximate the unitary used in phase estimation.  The best variant they reported was called split-operator with a fast fermionic Fourier transform~\cite{VerstraeteFFFT}, abbreviated as SO-FFFT. When investigating quantities such as the energy density, which converge towards some infinite lattice value, it is crucial to have fine control over lattice size to understand the rate of convergence and related finite size effects.  Unfortunately, SO-FFFT only works for lattices with length and width of the form $L=2^k$ for integer $k$.  While Kivlichan \textit{et al}~\cite{kivlichan2020improved} also explored approaches suitable for any lattice size, the resource costs were around an order of magnitude larger than SO-FFFT. 

More modern techniques than Trotterization have dramatically improved the asymptotic scaling with respect to accuracy using linear combinations of unitaries~\cite{childs2012hamiltonian} methods such as truncated Taylor series~\cite{Berry15,Babbush2016,meister2020tailoring} and qubitization~\cite{Low2019hamiltonian,QSP17,Poulin18,Babbush18}.  However, for relative errors in the range $0.5 \% - 0.05 \%$, the total permitted error is quite large so that constant factors are more important than asymptotic scaling.  In this regime, it is a game of constant factors, and so far refinements of second-order Trotterization have been more fruitful than development of more sophisticated algorithms. 

In this work, we both tighten the resource analysis of previous second-order Trotter approaches including SO-FFFT and introduce a new method that we called plaquette Trotterization.  We begin by analysing the Trotter error per step starting from a commutator bound~\cite{wecker2014gate,kivlichan2020improved,childs2019theory}.  Exactly evaluating the commutator bound is itself very difficult. Typically, commutator bounds are further loosened into a sum over many nested commutators between local operators (achieved by liberal use of the triangle inequality), which are enumerated computationally or by hand.  Instead, we present a tighter, novel approach to evaluating the commutator bound in terms of free-fermionic Hamiltonians that can then be efficiently evaluated.  This accounts for a large portion of our observed resource savings over the SO-FFFT estimates of Kivlichan \textit{et al}~\cite{kivlichan2020improved}. We further discuss the synthesis of $Z$-axis rotations into $T$-gates and the trade-offs between space and time costs due to Hamming weight phasing.  We find that for phase estimation of the TDL ground state density, our improvements gives $5.5\times$ to $9 \times$ reductions in the number of $T$-gates when compared against SO-FFFT on the $8 \times 8$ and $16 \times 16$ lattices.  Plaquette Trotterization can be used at all other lattices sizes where previous resource estimates were significantly higher.

\etc{Other variants of Trotter exist that we do not quantitatively study here.  Higher order Trotter has asymptotically better scaling than second order Trotter, but worse constant factors.  Furthermore, the complexity of evaluating the commutator bound increases rapidly with the Trotter order, so exact performance estimates are difficult, perhaps even intractable.  Nevertheless, since we are interested in a pre-asymptotic regime with large relative error, this favours second order Trotter.  Randomly compiled Trotter schemes such as qDRIFT~\cite{campbell2019random,ouyang2020compilation} can eliminate dependence on the number of Hamiltonian terms at the cost of worse scaling with respect to other parameters.  This can be especially attractive for $N$-qubit quantum chemistry Hamiltonians with $O(N^4)$ terms~\cite{mcardle2020quantum}, but Hubbard model Hamiltonians are very sparse with only $O(N)$ local terms, and only 2 Hamiltonian terms when grouped into interaction and hopping terms.  For such sparse Hamiltonians, qDRIFT will typically perform worse than standard second order Trotter.}

\section{The Hubbard Model Hamiltonian}

We consider Hamiltonians $H=H_h+ H_I$ with hopping terms of the form
\begin{equation}  \label{FullHop}
 H_h= \sum_{\sigma \in \uparrow, \downarrow}  \sum_{i \neq j} R_{i,j} a^{\dagger}_{i,\sigma} a_{j,\sigma}  ,
\end{equation}
where $\sigma$ labels the spin and Hermiticity entails that $R_{i,j}=R_{j,i}^*$ and we set $R_{i,i}=0$.   This part of the Hamiltonian is free fermionic and therefore exactly solvable.  Here we quickly review the most salient facts, but refer the reader to App.~\ref{AppFreeFermionic} for a longer discussion. We can find a unitary $U$ that diagonalizes $R$, so the eigenvalues of $R$ give the energy of the canonical fermionic modes of the Hamiltonian.  Due to $R_{i,i}=0$, the spectrum is symmetric about zero and the maximum (minimum) energies of $H_h$ correspond to summing over all the positive (negative) energy fermionic modes. This leads to $||H_h|| =  || R ||_1$ where $|| \ldots ||$ is the operator norm and $|| \ldots ||_1$ is the Schatten 1-norm or trace norm.  

\etc{Before we define our form for $H_I$, we define the more commonly encountered unshifted form
\begin{equation}
    \tilde{H}_{I} = \sum_i u \hat{n}_{i, \uparrow}  \hat{n}_{i, \downarrow}
\end{equation}
where $i$ indexes lattice sites and $\hat{n}=a^\dagger a$ is the number operator and $u$ is the interaction strength.  Shifting the chemical potential of the interaction Hamiltonian and adding the identity operator, we obtain the shifted interaction Hamiltonian
\begin{align}
 H_I & = \tilde{H}_{I}  -\frac{u}{2} \hat{N} + \frac{u}{4} \id ,
\end{align}
where $\hat{N}$ is the total electron number operator
\begin{equation}
\hat{N} =   \sum_{i} ( \hat{n}_{\uparrow, i} + \hat{n}_{\uparrow, i}) .
\end{equation}
Since the chemical potential shift $(u/2) \hat{N}$ and identity shift $(u/4)\id$ both commute with $H_h$ and $\tilde{H}_{I}$, we know the shifted and unshifted Hamiltonians commute and share an energy eigenbasis composed of eigenstates of $\hat{N}$.  Given such an eigenstate $\ket{\psi}$ with $\eta$ electrons (so that  $\hat{N} \ket{\psi} = \eta  \ket{\psi}$), then from $(H_h + \tilde{H}_I)\ket{\psi}=\tilde{E}\ket{\psi}$ we know  $(H_h + H_I)\ket{\psi} = E\ket{\psi} $ with $E=\tilde{E}- u \eta / 2 + u/4$.  Therefore,  within an $\eta$-electron subspace, shifting the Hamiltonian changes the spectrum by a known additive constant that can be removed by classical processing. We will see that shifting the chemical potential both tightens Trotter error bounds and reduces gate count per Trotter step. Herein, we write our interaction Hamiltonian as
\begin{align}
 & H_I = u \sum_{ i }  (\hat{n}_{i, \uparrow} - \id/2)( \hat{n}_{i, \downarrow}-\id/2) \\
 & =   \frac{u}{4} \sum_{ i } \hat{z}_{i,\uparrow}\hat{z}_{i,\downarrow},
\end{align} 
where we introduced the shorthand $\hat{z}:=2 \hat{n}-\id $ and assume throughout that $u>0$.  } 

Many of our results hold for a general Hubbard Hamiltonian, but we also focus on TDL where $R_{i,j}=\tau$ when $i$ and $j$ are nearest neighbours on a square lattice and otherwise $R_{i,j}=0$. For TDL, the hard simulation regime corresponds to $4 < u / \tau <  12$ and $10\%-20\%$ below half-filling ~\cite{zheng2017stripe}.
 
\setlength{\tabcolsep}{10pt}
\begin{table*}  \begin{tabular}{lllllll}  \toprule
 & $L$ & \textbf{4} & \textbf{6} & \textbf{8} & \textbf{12} & \textbf{16} \\ \midrule
 & $W_{\mathrm{SO}} \leq$ &  $1.4*10^3$ & - & $5.5*10^3$ & - & $2.2*10^4$ \\
\textbf{SO-FFFT} & $N_T$ &  $2.6*10^2$ & - & $1.7*10^3$ & - & $1.0*10^4$ \\
(from~\cite{kivlichan2020improved})  & $N_R$ & $6.8*10$ & - & $2.9*10^2$ & - & $1.2*10^3$ \\ \midrule
 & $W_{\mathrm{SO}} \leq$ & $8.7*10$ & - & $3.5*10^2$ & - & $1.4*10^3$ \\
\textbf{SO-FFFT+} & $N_T$ & $2.6*10^2$ & - & $1.7*10^3$ & - & $1.0*10^4$ \\
 & $N_R$ & $3.6*10$ & - & $1.6*10^2$ & - & $7.1*10^2$ \\ \midrule
 & $W_{\mathrm{PLAQ}}\leq$ & $1.3*10^2$ & $3.0*10^2$ & $5.3*10^2$ & $1.2*10^3$ & $2.1*10^3$ \\
\textbf{PLAQ} & $N_T$ & $1.9*10^2$ & $4.3*10^2$ & $7.7*10^2$ & $1.7*10^3$ & $3.1*10^3$ \\
& $N_R$ & $6.4*10$ & $1.4*10^2$ & $2.6*10^2$ & $5.8*10^2$ & $1.0*10^3$ \\ \bottomrule
 \end{tabular} 
 \caption{Rigorous upperbounds on the error constant $W$ (total error $Ws^3$ for simulating $e^{i H s}$) and non-Clifford ($N_T$ is the number of $T$-gates and $N_R$ is the number of arbitrary angle $Z$-axis rotations) costs of implementing a single Trotter step of the TDL Hubbard model on a $L \times L$ lattice with $u/\tau=4$.  We compare previous SO-FFFT results~\cite{kivlichan2020improved} with our improved SO-FFFT+ analysis for the allowable lattice sizes $L=2^k$.  We also show results for PLAQ, we highlight that it works on other lattice sizes $L \neq 2^k$ with some examples.} \label{Table_Compare1}
\end{table*}

\begin{figure}[t!]
    \centering
    \includegraphics[width=200pt]{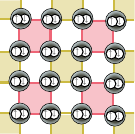}
    \caption{The interaction graph for a $4 \times 4$ Hubbard model on a square periodic lattice. Each site contains 2 fermions, one for each spin direction. For PLAQ-Trotterization, we divide the lattice into disjoint pink and gold plaquettes, such that each edge belongs to only 1 plaquette.}
    \label{fig:PLAW}
\end{figure}

\section{Improving the split-operator algorithm}

The second order Trotter formulae takes an ordered decomposition of the Hamiltonian $H=\sum_{j=1}^{m} H_j$ and approximates 
\begin{equation}
e^{i H  s} \approx \prod _{j=1}^m e^{i H_j s/2}  \prod _{j=m}^1 e^{i H_j s/2}  ,  
\end{equation}
where the second product runs in reverse index ordering.  While there is a product of $2m$ unitaries, this can be reduced to $2(m-1)$ by merging the middle two exponentials and the first/last exponentials.  Given such a decomposition, the error in the Trotter step is bounded by $\epsilon \leq W s^3$ where $W$ is a constant established by the well-known commutator bounds~\cite{wecker2014gate,kivlichan2020improved,childs2019theory}
\begin{align} \label{CommBoundGeneral}
    W = & \frac{1}{12} \sum_{b=1,2} \left| \left| \nonumber \sum_{c>b, a>b} [[ H_b , H_c ], H_a ] \right|\right|  
    \\ & + \frac{1}{24} \sum_{b=1,2} \left|\left| \sum_{c>b} [[ H_b , H_c ], H_b] \right|\right| .
\end{align}
The value of $W$ is highly dependent on our choice of Trotterization decomposition and can be difficult to evaluate. The split-operator (SO) approach uses two terms $H_1=H_I$ and $H_2=H_h$ (or interchanged).  However, to be implemented on a quantum computer $e^{i s H_h  }$ requires further decomposition. For any hopping Hamiltonian, it is possible to exactly diagonalize $e^{i s H_h  }$ using a product of $O(L^4)$ Givens rotations~\cite{wecker2015progress,kivlichan2018quantum,jiang2018quantum}, and while this is formally efficient, $O(L^4)$ is very large in practice.  For the special case of a periodic square $L \times L$ lattice with $L=2^k$ for integer $k$, we can instead use a fast-fermionic Fourier transform~\cite{VerstraeteFFFT} (FFFT) to diagonalize $e^{i s H_h  }$ and we refer to this approach as SO-FFFT. It was found by Kivlichan \textit{et al.}~\cite{kivlichan2020improved} that SO-FFFT achieves gate counts that are orders of magnitude better than using Givens rotations or any other approach that they considered.  

We improve on the prior analysis of SO-FFFT in two regards and refer to our improved results as SO-FFFT+.  First, we reduce the gate complexity of implementing $e^{i s H_I}$ by exploiting our choice of chemical shift. Using the standard form of $H_I$, each term $\hat{n}_{i, \uparrow}  \hat{n}_{i, \downarrow}$ transforms under Jordan-Wigner to 3 non-trivial Pauli operators, and so $e^{i s H_I}$ requires $3 L^2$ $Z$-axis rotations.  Using our shifted Hamiltonian, each  $\hat{z}_{i,\uparrow}\hat{z}_{i,\downarrow}$ term Jordan-Wigner transforms to a single Pauli operator of the form $Z \otimes Z$, so we only need $L^2$ arbitrary $Z$-axis rotations to realise $e^{i s H_I}$.  Some $Z$-axis rotations are also required to realised $e^{i s H_h}$, and our resulting resource savings are summarised in Table.~\ref{Table_Compare1}.  Second, we greatly improve the Trotter error bound $W_{SO}$ for split-operator methods.  Kivlichan \textit{et al.}~\cite{kivlichan2020improved}  bounded $W_{\mathrm{SO}}$ by expanding the commutator bounds in terms of Pauli operators, liberally applying the triangle inequality and computationally counting the number of non-zero terms.  However, each application of the triangle inequality dramatically loosens the bound.  Whereas, for the TDL we prove that
$W_{SO} \leq \mathrm{min}[W_{SO1}, W_{SO2}]$ where
\begin{align} \label{Wbounds1}
W_{SO1} & \leq  \left( \frac{u^2 }{12} \right) || H_h || + \left( \frac{u \tau^2}{12} \right) L^2(\sqrt{5}+8)  , \\ \label{Wbounds2} 
W_{SO2} & \leq \left( \frac{u \tau^2}{6} \right) L^2(\sqrt{5}+8) + \left( \frac{u^2 }{24}\right)  || H_h ||  ,
\end{align}
where $ || H_h ||= ||R||_1 $ can be efficiently computed because $H_h$ is free-fermionic. The minimisation over $W_{SO1}$ and $W_{SO2}$ is due to the two possible ordering for SO.  We give an example of the resulting improvements in Table~\ref{Table_Compare1}, where the reductions average $16\times$. The proof has two main steps. First, bounding $\| [[ H_I , H_h ], H_I ] \|$ and $\| [[ H_I , H_h ], H_h ] \|$, which is the most technical step.  Then combining these bounds to obtain $W_{SO1}$ and $W_{SO2}$ is straightforward (see App.~\ref{SquareTheorem} for details).  Each step of the proof will introduce a lemma that is applicable to any Hubbard Hamiltonian, so the results can be easily extended beyond TDL.

\textit{First Lemma and proof sketch.-} Our first bound is the especially elegant result that for any Hubbard Hamiltonian
\begin{equation} \label{FirstBound}
  \|   [[ H_I ,H_h   ], H_I ] \|   \leq u^2  \|H_h  \| .
\end{equation}
We provide a detailed proof in App.~\ref{AppFirstLemma} and sketch the main steps here.  Using the anti-commutation relations of fermionic operators, we have
\begin{equation}
 [[ H_I ,H_h   ], H_I ] = -\frac{u^2}{2} \sum_{i \neq j} \hat{z}_{i, \uparrow} \hat{z}_{i, \downarrow}\hat{z}_{j, \uparrow} \hat{z}_{j, \downarrow}  B_{i,j} -\frac{u^2}{2} H_h  \nonumber ,
\end{equation}
where $B_{i,j}=\sum_{\sigma}R_{i,j} a^{\dagger}_{i,\sigma} a_{j,\sigma}$.  We will construct a unitary transformation to simplify the first set of terms. We define the unitary $V_j := (\id +i  \hat{z}_{j, \uparrow} \hat{z}_{j, \downarrow})/\sqrt{2}$ and use anti-commutation relations to verify that
\begin{equation}
 (V_i \otimes V_j) B_{i,j}(V_i \otimes V_j)^\dagger   = - (\hat{z}_{i, \uparrow} \hat{z}_{i, \downarrow})(\hat{z}_{j, \uparrow} \hat{z}_{j, \downarrow}) B_{i,j} .
\end{equation}
Furthermore, for $k \neq i,j$ the unitary $V_k$ commutes with $B_{i,j}$ so we can consider the unitary over all sites 
\begin{equation}
    \bar{V} := \prod_{j=1}^{n} V_j ,
\end{equation}
and we have
\begin{equation}
    \bar{V} B_{i,j} \bar{V}^\dagger   = - (\hat{z}_{i, \uparrow} \hat{z}_{i, \downarrow})(\hat{z}_{j, \uparrow} \hat{z}_{j, \downarrow}) B_{i,j} .
\end{equation}
Summing over $i \neq j$, gives
\begin{equation}
 \bar{V} H_h \bar{V}^\dagger   = - \sum_{i \neq j} \hat{z}_{i, \uparrow} \hat{z}_{i, \downarrow}\hat{z}_{j, \uparrow} \hat{z}_{j, \downarrow} B_{i,j} ,
\end{equation}
and so $ [[ H_I ,H_h   ], H_I ]=(u^2/2)(\bar{V} H_h \bar{V}^\dagger -  H_h) $.  Taking the operator norm, making a single application of the triangle inequality and using unitary invariance of the norm we deduce Eq.~\eqref{FirstBound}.

\textit{Second Lemma and proof sketch.-} For any Hubbard Hamiltonian we have
\begin{equation} \label{SecondLemmaEq}
    \|   [[ H_I , H_h   ], H_h ] \| \leq  \frac{u}{2} \sum_i \left( \|  [T_i,H_h  ]  \|  + 2 \| T_i \|^2 \right) ,
\end{equation}
where $T_i=\sum_{j} (B_{i,j}+B_{i,j}^\dagger)/2$.  Since $T_i$ is free-fermionic we can efficiently evaluate $\| T_i \|^2$.  Furthermore, the commutator of two free-fermionic Hamiltonians, is again free fermionic so each $\|  [T_i,H_h  ]  \|$ is efficiently computable.   A detailed proof is given in App.~\ref{AppSecondLemma} and here we outline the main steps.

The first step in proving this lemma is an application of the triangle inequality, so that
\begin{equation} 
    \|   [[ H_I ,H_h   ], H_h ] \| \leq \frac{u}{4} \sum_{i}  \|   [[ z_{i, \uparrow }z_{i, \downarrow } ,H_h   ], H_h ] \| ,
\end{equation}
Next we use anti-commutation relations to find $[ z_{i, \uparrow }z_{i, \downarrow } ,H_h   ]=2z_{i, \uparrow }z_{i, \downarrow } T_i$ .  Nesting the commutator, we have  $[[ z_{i, \uparrow }z_{i, \downarrow } ,H_h   ],H_h ]=2[z_{i, \uparrow }z_{i, \downarrow } T_i, H_h]$. Next, we use the commutator identity $[PQ,R]=P[Q,R]+[P,R]Q$ and $[ z_{i, \uparrow }z_{i, \downarrow } ,H_h   ]=2z_{i, \uparrow }z_{i, \downarrow } T_i$ to conclude that 
\begin{equation}
    [[ z_{i, \uparrow }z_{i, \downarrow } ,H_h   ],H_h ]=2 z_{i, \uparrow }z_{i, \downarrow } [ T_i, H_h] + 4  z_{i, \uparrow }z_{i, \downarrow } T_i^2 .
\end{equation}
Taking the operator norm, applying the triangle inequality, using $\| z_{i, \uparrow }z_{i, \downarrow } \|=1$, and summing over all indices $i$ gives Eq.~\eqref{SecondLemmaEq}.  
For the special case of TDL, one finds that whenever $L \geq 4$ these take constant values $\| T_i \|=4 \tau$ and $\| [T_i,H_h  ] \|= 4 \sqrt{5} \tau^2$.  Since there are $L^2$ values of $i$, the summation gives
\begin{equation} 
    \|   [[ H_I , H_h   ], H_h ] \| \leq  u \tau^2  \left( 2 \sqrt{5}  +  16 \right) L^2 ,
\end{equation}
which leads to one term in each of Eq.~\eqref{Wbounds1} and Eq.~\eqref{Wbounds2}.

\section{Plaquette Trotterization}

In addition to improving the analysis of SO-FFFT, we introduce a different second-order Trotter layering that we call plaquette Trotterization or PLAQ for short. A significant shortcoming of SO-FFFT is that the FFFT only works for lattices of size $L=2^k$ and the FFFT requires a substantial number of non-Cliffords to perform. We partition the edges of the square lattices into two colours, pink and gold, such that each edge belongs to a single plaquette (as shown in Fig.~\ref{fig:PLAW}) and we partition the Hamiltonian as $H_h=H_h^p+H_h^g$ where the $H_h^p$ ($H_h^g$) contains all the interactions corresponding to pink (gold) edges. Such a partition is possible whenever $L$ is even, though PLAQ can be easily modified to also work for odd $L$ by having a small number of plaquettes with only 1 or 2 edges.  Setting $H_1=H_I$, $H_2=H_h^p$ and $H_3=H_h^g$ and using the general bound of Eq.~\eqref{CommBoundGeneral} we obtain a bound containing  $W_{SO2}$ of Eq.~\eqref{Wbounds2} and some additional terms.  Noting that $||    [[ H_h^p , H_h^g ], H_h^g ]   || =||   [[ H_h^p , H_h^g  ], H_h^p ]    ||$ the additional terms can be combined to obtain
\begin{equation}
    W_{PLAQ} \leq  W_{SO2} + \frac{3}{24} \|    [[ H_h^p , H_h^g ], H_h^g ]   \| .
\end{equation}
Since all the additional terms involved are free-fermionic these can be efficiently computed (for details see App~\ref{AppSecondLemma} and Table~\ref{HhopValues} therein). Table~\ref{Table_Compare1} shows that this leads to only a slight increase for PLAQ error bounds relative to SO-FFFT+. To implement unitaries of the form $e^{i s H_h^p}$ requires some further decomposition. We note that  
\begin{equation}
 H_h^p=\sum_{k, \sigma} H_h^{p(k, \sigma)} ,
\end{equation}
where $H_h^{p(k, \sigma)}$ is a plaquette Hamiltonian acting on the 4 hopping terms corresponding to the $k^{\mathrm{th}}$ plaquette in the pink set and with spin index $\sigma$.   Each of these individual plaquettes commute so, 
\begin{equation}
e^{i s H_h^p}=\prod_{k, \sigma}e^{i s H_h^{p(k, \sigma)}} ,
\end{equation}
exactly and without further Trotter error.  Each plaquette Hamiltonian acts on 4 fermions, which can be implemented with 8 $T$ gates and 2  $Z$-axis rotations of equal magnitude (see App~\ref{App:realisePLAQ} for details).   Counting $L^2/4$ plaquettes of colour pink and 2 spin sectors, $e^{i s H_h^{p}/2}$ requires  $4L^2$ $T$-gates and $L^2$  $Z$-axis rotations.  The full Trotter step uses 2 implementations of $e^{i s H_h^{p}/2}$ and one of $e^{i s H_h^{g}}$, which totals to $12 L^2$ $T$-gates and $3 L^2$ $Z$-axis rotations. Adding a cost of $L^2/2$ $Z$-axis rotations for $e^{i s H_I}$, implemented just as in SO-FFFT+, we arrive at a total of $12 L^2$ $T$-gates and $4 L^2$ arbitrary $Z$ rotations per Trotter step.

\begin{figure}[t!]
    \centering
    \includegraphics{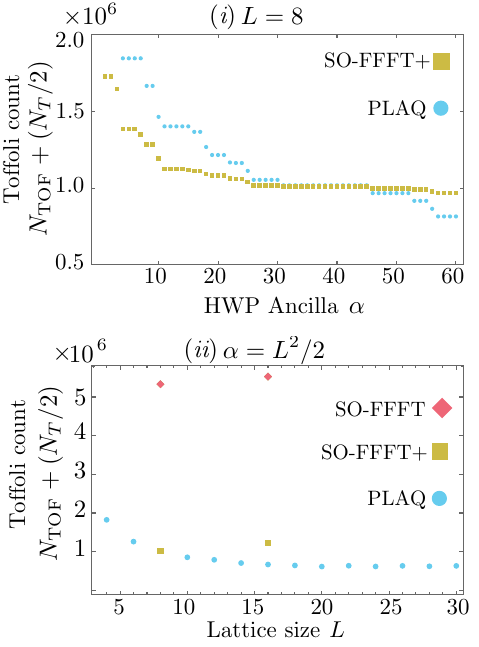}
    \caption{Upper bounds on the total Toffoli count (all $T$-gates catalyzed using Toffoli states) for phase estimation of the TDL with $u=4$, $\tau=1$ and additive error $\epsilon=0.0051 L^2$ (approx. half a percent of the total system energy~\cite{kivlichan2020improved,HalfPercent}).  In (\textit{i}) we fix $L=8$ and compare PLAQ and SO-FFFT+ allowing for a varying number of logical ancilla $\alpha$ used for Hamming weight phasing.  In (\textit{ii}), we fix the number of logical ancilla to $L^2/2$ and vary $L$.  Also shown in (\textit{ii}) are the previous data points for SO-FFFT taken from Table 1 of Kivlichan \textit{et al.}~\cite{kivlichan2020improved}.  }
    \label{fig:Tcounts}
\end{figure}

\setlength{\tabcolsep}{5pt}
\begin{table}[t]
    \centering
    \begin{tabular} {c c c c c} \toprule
& \text{Lattice size}   &  logical qubits &  TOF gates & T gates \\
$u/\tau$ & $L$ & $N_Q$ & $N_{\mathrm{TOF}}$ &  $N_{T}$ \\ \midrule
4 & 8 & 162 & $1.8*10^5$ & $1.7 * 10^6$ \\
4 & 10 & 252 & $1.8*10^5$ & $1.3*10^6$ \\
4 & 12 & 362 & $1.9*10^5$ & $1.2* 10^6$ \\
4 & 14 & 492 & $1.9*10^5$ & $1.0* 10^6$ \\
4 & 16 & 642 & $1.9*10^5$ & $9.5*10^5$ \\
4 & 18 & 812 & $1.9*10^5$ & $9.0*10^5$ \\
4 & 20 & 1002 & $2.0*10^5$ & $8.4*10^5$ \\
4 & 22 & 1212 & $1.9*10^5$ & $8.9*10^5$ \\
4 & 24 & 1442 & $1.9*10^5$ & $8.5*10^5$ \\
4 & 26 & 1692 & $1.9*10^5$ & $8.8*10^5$ \\
4 & 28 & 1962 & $2.0*10^5$ & $8.5*10^5$ \\
4 & 30 & 2252 & $2.0*10^5$ & $8.7*10^5$ \\
4 & 32 & 2562 & $2.0*10^5$ & $8.7*10^5$ \\ \midrule
8 &  8 & 162 & $4.3* 10^5$ & $4.1* 10^6$ \\
8 & 10 & 252 & $4.4* 10^5$ & $3.3* 10^6$ \\
8 & 12 & 362 & $4.6* 10^5$ & $2.9* 10^6$ \\
8 & 14 & 492 & $4.6* 10^5$ & $2.5* 10^6$ \\
8 & 16 & 642 & $4.6* 10^5$ & $2.3* 10^6$ \\
8 & 18 & 812 & $4.6* 10^5$ & $2.2* 10^6$ \\
8 & 20 & 1002 & $4.7* 10^5$ & $2.0* 10^6$ \\
8 & 22 & 1212 & $4.6* 10^5$ & $2.2* 10^6$ \\
8 & 24 & 1442 & $4.7* 10^5$ & $2.1* 10^6$ \\
8 & 26 & 1692 & $4.6* 10^5$ & $2.1* 10^6$ \\
8 & 28 & 1962 & $4.6* 10^5$ & $2.1* 10^6$ \\
8 & 30 & 2252 & $4.7* 10^5$ & $2.1* 10^6$ \\
8 & 32 & 2562 & $4.7* 10^5$ & $2.1* 10^6$ \\
 \bottomrule
    \end{tabular}
    \caption{Resources estimates using PLAQ to perform phase estimation of the TDL with $u/ \tau=4$ and $u / \tau =8$ and error of approximately half a percent of the total system energy~\cite{kivlichan2020improved,HalfPercent}). The total number of logical qubits is $N_Q= 2L^2 + \alpha + 2$ with $\alpha= L^{2}/2$ qubits used for Hamming weight phasing.  The $+2$ refers to 1 qubit used for phase estimation and 1 qubit used for repeat until success synthesis~\cite{bocharov15b}.  If $T$ gates are performed by catalysis then an additional qubits is needed.  Within a fault-tolerant quantum computer, $N_Q$ refers to the number of logical qubits.  To obtain a full architecture-specific resource estimate one must also include the overhead of quantum error correction, the footprint for any magic state distillation factories, any routing overheads or storage of a $T$-catalyst.}
    \label{tab:SynthesizedResource}
\end{table}

\section{Rotation synthesis}

When $L=2^k$, we can compare the gate counts per Trotter step of SO-FFFT+ and PLAQ.  The total $T$-gates plus Pauli rotations for SO-FFFT+ is larger than for PLAQ, and when $L=16$ this difference is over a factor $2.7 \times$.  However, in a fault-tolerant device arbitrary angle $Z$-axis rotations require further synthesis, and using standard techniques~\cite{bocharov15b,kliuchnikov13,gosset14,bocharov15,RS14} this is typically in the range of $10-50$ $T$-gates per $Z$-axis rotation.  Such counting suggests SO-FFFT+ has a lower $T$-count than PLAQ on applicable lattice sizes. However, if we use a more sophisticated synthesis strategy described below, then PLAQ achieves a lower gate count per Trotter step.

When a collection of $Z$-axis rotations have equal angle (up to a unimportant sign), they can be more efficiently synthesized using Hamming weight phasing (HWP)~\cite{gidney2018halving,kivlichan2020improved}. The basic idea of HWP, is that the rotation $\otimes_j^m e^{i \theta Z_j}$ acting on $m$ qubits in computational basis state $\ket{x}$ will impart a phase $e^{i \theta (2w(x)-m) }$ where $w(x)=\sum_j x_j$ is the Hamming weight of bit-string $x$.  Using an ancillary register of size $\log_2(m)$ and computing a binary representation of the Hamming weight of $x$, so $\ket{x} \rightarrow \ket{x}\ket{ w(x) }$ we can then apply $\left \lceil{ \log_2(m)+1}\right \rceil$ $Z$-axis rotations to the ancillary register in the state $\ket{ w(x) }$ in order to acquire the correct phase and then uncompute the Hamming weight calculation.  This exponentially reduces the number of $Z$-axis rotations required, but requires $\alpha$ additional ancilla and $\alpha$ Toffoli gates (equivalently $4 \alpha$ $T$-gates~\cite{NovelToffoli}). Gidney showed~\cite{gidney2018halving,kivlichan2020improved} that when $m=2^k$ we have $\alpha=m-1$ and that $\alpha \leq m-1$ for general $m$. In the limit of large $m$, an arbitrary rotation now costs closer to 4 $T$-gates. We present more details on HWP in App~\ref{App:HWP}. We remark that by combining the results of Ref.~\cite{meuli2019role} and Ref.~\cite{boyar2005exact} one can tighten Gidney's bounds to show the Hamming weight can be computed with the cost determined by $\alpha=m-w(m)$ where $w(m)$ is the Hamming weight of the binary representation of integer $m$. 

PLAQ is especially well suited to exploit HWP because in $e^{i s H_I^{p}}$ or $e^{i s H_I^{g}}$, every $Z$-axis rotation is of equal angle.  In contrast, SO-FFFT+ can only partly exploit HWP because the rotation angles depends on the eigenvalues of $R$ (recall Eq.~\ref{FullHop}). While the matrix $R$ has some degeneracy that can be exploited, it can not exploit HWP to the same, comprehensive extent as PLAQ. Therefore, given the numbers in Table~\ref{Table_Compare1}, we expect that HWP can lend an advantage to PLAQ over SO-FFFT+ (for $L=2^k$ lattices) and in the next section we find that this is indeed the case.  

\section{Phase estimation}

Next, we consider using the phase estimation algorithm using the same assumptions and parameters used in Kivlichan \textit{et al.}~\cite{kivlichan2020improved}.  \etc{In particular, we ideally wish to perform phase estimation on the unitary $ \exp(i H t)$.  However, due to Trotter errors we instead perform some other unitary $U$ that approximates $\exp(i H t)$.  By unitarity of $U$, we can always find an effective Hamiltonian $H_{\mathrm{eff}}$ such that $U= \exp(i H_{\mathrm{eff}} t)$. Furthermore, from App E of  Ref.~\cite{kivlichan2020improved}, we know that $H_{\mathrm{eff}}$ can be chosen so that $\| H - H_{\mathrm{eff}} \| \leq W t^2 $ (assuming the regime that $Wt^3 \ll 1$, which we always numerically verify) and where $W$ is the Trotter error constant bounded earlier.  Whatever value of $W$ we obtain, we can always reduce the error by decreasing $t$. However, smaller $t$ entails more bits of precision are needed in phase estimation, leading to more uses of $U$ and higher gate complexity.  Indeed, optimising the value of $t$, one finds that the optimal gate complexity is proportional to $\sqrt{W}$ (see App.~\ref{AppPhaseEstimation} for details)}. Assuming preparation of an approximate ground state with fidelity $f$ overlap with the true ground state, then an ideal implementation of phase estimation would with probability $f$ return the ground state energy (up to additive error $W t^2$). How to efficiently prepare states with sufficiently good overlap with the ground state is an important and non-trivial problem that we will not address here.

Our results for PLAQ are presented in Table \ref{tab:SynthesizedResource} and the resource counting methodology is described in more detail in App.~\ref{AppPhaseEstimation}. In Fig.~\ref{fig:Tcounts}, we convert all gates to Toffoli gates and make a comparison between PLAQ and spilt-operator estimates. We choose to work in units of Toffoli gates instead of $T$-gates since recent, low-overhead magic state distillation factories output Toffoli states~\cite{gidney2019efficient,AWSarchitecture}. The Toffoli count for PLAQ can be reproduced from Table~\ref{tab:SynthesizedResource} by assuming that 2 $T$-gates can be performed using 1 Toffoli gate (given access to a catalyst~\cite{gidney2018halving,gidney2019efficient,beverland2020lower}) so that the total Toffoli count is $N_{\mathrm{TOF}}+(N_{T}/2)$.
 
 Fig.~(\ref{fig:Tcounts}\textit{i}) illustrates the trade-off between gate count and the number of ancilla used for HWP. We see that when few ancilla are available to support HWP, then SO-FFFT+ has lower $T$-count than PLAQ.  However, PLAQ is better able to exploit HWP, so as we increase the available ancillas it outperforms SO-FFFT+. 
 
 In Fig.~(\ref{fig:Tcounts}\textit{ii}) we allocate a constant fraction of all qubits to HWP and vary $L$, observing comparable performance between PLAQ and SO-FFFT+.  Crucially, PLAQ improves on the best previous $T$-counts of SO-FFFT by a factor $5.5\times$ (at $L=8$) and a factor $7.8\times$ (at $L=16$) and provides this performance across all even $L$ rather than the sparse data-points provided by the split-operator method. Recall that Givens rotations and fermionic swap networks can operate at any $L$ value, but they have orders of magnitude higher Toffoli counts and would not be visible on the scale of our plots. A curious feature is that the Toffoli count reduces with $L$, which is a consequence of allowing for a relative error that grows with $L^2$.  The growth in the allowed error leads to a reduction in the number of Trotter steps needed, which almost exactly matched the $O(L^2)$ gate-count per Trotter step.  In a fault-tolerant device, a factor $r$ reduction in $T$-count will lead to a factor $r$ reduction in runtime and consequently also slight reduction in the number of physical qubits per logical qubit. The physical qubit requirements and algorithm runtime for specific architectures is discussed further in the companion paper Ref.~\cite{AWSarchitecture}

\section{Conclusions}

The Hubbard model was already identified as an interesting physical system to simulate on an early fault-tolerant quantum computer. We have further improved these prospects by tightening error bounds and reducing gate complexity, so that now there is a potential demonstration of a useful quantum advantage requiring fewer than 1 million Toffoli gates. To the best of our knowledge, this is the lowest Toffoli count amongst rigorously studied quantum algorithms outside the reach of current classical computers.  This makes implementing this algorithm a compelling milestone goal for fault-tolerant quantum computers.   

\textit{Acknowledgements.-} We thank Fernando Brand\~ao for proposing a study of the Hubbard model and useful early discussions. We thank Yuan Su for discussions on commutator bounds and Sam McArdle for detailed feedback on the manuscript.  


%

\appendix

\section{Free fermionic Hamiltonians}
\label{AppFreeFermionic}

Here, we review some basic facts about fermionic Hamiltonians.  We say an operator is free fermionic if it can be decomposed as
\begin{equation}
    H = \sum_{i,j} Q_{i,j} a^\dagger _i a_j .
\end{equation}
We see $H$ is Hermitian if $Q$ is Hermitian.  Since $Q$ is $n$ by $n$ for an $n$ fermion system, we can efficiently diagonalize Hermitian $Q$ and find $Q=U^\dagger D U$ with real diagonal $D$.  Here $U$ are unitary operators on the mode space, so that 
\begin{align}
    H   = \sum_{k} D_{k} b^\dagger_k b_k ,
\end{align}
where
\begin{align}
    b_k := \sum_{j} U_{j,k} a_j ,
\end{align}
are the canonical modes of the system and $D_k$ are their associated energies.  

From this we can determine the spectrum of $H$, which has the form
\begin{equation}
    \lambda_{S} = \sum_{k \in S} D_k ,
\end{equation}
where $S \subseteq [1,n]$ is a set denoting which fermion modes are filled.  Clearly, the ground state corresponds to filling all the lowest energy states, so
\begin{align}
   \mathrm{min}_S \{   \lambda_{S} \} & = \sum_{k :  D_k < 0} D_k , \\ \nonumber
    \mathrm{max}_S \{   \lambda_{S} \} &= \sum_{k :  D_k > 0} D_k , 
\end{align}
and the operator norm of $H$ is the maximum of these two numbers. In the special case that $H$ is a hopping Hamiltonian, so that $Q_{j,j}=0$, then
\begin{equation}
 \mathrm{min}_S \{   \lambda_{S} \} = -  \mathrm{max}_S \{   \lambda_{S}  \}
\end{equation}
so
\begin{equation}
    ||H||= \frac{1}{2} \sum_k |D_k| =\frac{1}{2} ||Q||_1 ,
\end{equation}
where $|| \ldots ||_1$ is the Schatten 1-norm.  In the main text, we consider Hubbard models with two spin sectors with symmetric hopping Hamiltonians, so that
\begin{equation} \label{SubBlock}
    Q= \left( \begin{array}{cc}
    R & 0 \\ 
    0 & R \end{array} \right) ,
\end{equation}
in which case it is convenient to work with $R$ for a single spin sector, and we have
\begin{equation} \label{Benergy}
    ||H|| = || R ||_1.
\end{equation}
The anticommutation relations state that
\begin{align}
    a^\dagger _j a_k + a_k a_{j}^\dagger & = \delta_{j,k} ,  \\ \nonumber
     a^\dagger_j a_k^\dagger + a_k^\dagger a_{j}^\dagger & = 0 , \\ \nonumber
      a_{j} a_k + a_k a_{j} & = 0  .
\end{align}
From these we can derive the commutator relation
\begin{align}
   [ a^\dagger _j a_k , a_l^\dagger a_m ] & = \delta_{k,l} a_{j}^\dagger a_{m} -\delta_{j,m} a^\dagger_l a_k  .
\end{align}
Given two free fermionic operators $H_Q$ and $H_P$ (not necessarily Hermitian) with associated matrix representations $Q$ and $P$, we have that
\begin{align}
    [ H_Q, H_P ]  = & \sum_{j,k,l,m} Q_{j,k} P_{l,m} [  a^\dagger _j a_k , a_l^\dagger a_m  ] \\ \nonumber
     =  & \sum_{j,k,l,m} Q_{j,k} P_{l,m} (\delta_{k,l} a_{j}^\dagger a_{m} -\delta_{j,m} a^\dagger_l a_k  ) \\ \nonumber
     = & \sum_{j,k,l,m} Q_{j,k} P_{l,m} \delta_{k,l} a_{j}^\dagger a_{m} \\ \nonumber
     & - \sum_{j,k,l,m} Q_{j,k} P_{l,m} \delta_{j,m} a^\dagger_l a_k   \\ \nonumber
     = & \sum_{j,k,m} Q_{j,k} P_{k,m}  a_{j}^\dagger a_{m} - \sum_{j,k,l} Q_{j,k} P_{l,j}  a^\dagger_l a_k   \\ \nonumber
       = & \sum_{j,m} [Q P]_{j,m}  a_{j}^\dagger a_{m} - \sum_{k,l} [P Q]_{l,k}   a^\dagger_l a_k   ,
\end{align}
where $[ \ldots ]_{j,k}$ denotes the matrix element on the matrix contained inside the brackets.  Making a change of variables in the second summation $l \rightarrow j$ and $k \rightarrow m$, we get
\begin{align}
    [ H_Q, H_P ]  & = \sum_{j,m} [Q P - P Q ]_{j,m}  a_{j}^\dagger a_{m}  ,
\end{align}
and so this is a new free fermionic Hamiltonian the associated matrix representation given by $[Q, P]$. In the case of interest where Eq.~\eqref{SubBlock} holds, we can evaluate the commutator for a single sub-block.

\section{Anti-commutation}
\label{AntiCommute}

Here we show that operators of the form $\hat{z}_{k,\uparrow} \hat{z}_{k,\downarrow}$ and $B_{i,j}=\sum_{\sigma}R_{i,j} a^{\dagger}_{i,\sigma} a_{j,\sigma}$ form a Pauli-like algebra in that either commute or anti-commute. This will be used throughout subsequent appendices, for instance in Eq.~\eqref{ABresult}. Commutation when $k \neq  i,j$ is obvious so we focus on anti-commutation between $\hat{z}_{i,\uparrow} \hat{z}_{i,\downarrow}$ and $B_{i,j}$.  First we establish that
\begin{align}
    \{ \hat{z}_{i,\uparrow} \hat{z}_{i,\downarrow} , a_{i, \sigma}  \} & = 0  ,\\ \nonumber
    \{ \hat{z}_{i,\uparrow} \hat{z}_{i,\downarrow} , a^\dagger_{i, \sigma}  \} & = 0 ,
\end{align}
where $\sigma$ could be up or down. We have
\begin{align} \nonumber
    \{ \hat{z}_{i,\uparrow} \hat{z}_{i,\downarrow} , a_{i, \sigma}  \} & =  \{(2\hat{n}_{i, \bar{\sigma}} - \id ) (2\hat{n}_{i, \sigma} - \id ) , a_{i, \sigma}  \} \\ \nonumber
    & = (2\hat{n}_{i, \bar{\sigma}} - \id ) \{ (2\hat{n}_{i, \sigma} - \id ) , a_{i, \sigma}  \} \\ \label{AiAnnAntiComt}
 & =2 (2\hat{n}_{i, \bar{\sigma}} - \id )\left( \{ \hat{n}_{i, \sigma} , a_{i, \sigma}  \} -   a_{i, \sigma}  \right) ,
\end{align}
where in the second line we have used that $(2\hat{n}_{i, \bar{\sigma}} - \id)$ commutes with all other operators in the expression.  In the last line we have evaluated the anti-commutator for the identity term.  The anti-commutator $\{ \hat{n}_{i, \sigma} , a_{i, \sigma}  \}$ is
\begin{align} \label{NumAnnAntiCom}
    \{ \hat{n}_{i, \sigma} , a_{i, \sigma}  \} & =  a_{i, \sigma}^\dagger a_{i, \sigma}  a_{i, \sigma}   +    a_{i, \sigma} a_{i, \sigma}^\dagger a_{i, \sigma}  \\ \nonumber 
    & =     a_{i, \sigma} a_{i, \sigma}^\dagger a_{i, \sigma}  \\ \nonumber 
    & =     (1 - a^\dagger_{i, \sigma} a_{i, \sigma}) a_{i, \sigma}  \\ \nonumber 
    & =     a_{i, \sigma} - a^\dagger_{i, \sigma} a_{i, \sigma} a_{i, \sigma}   \\ \nonumber 
    & =  a_{i, \sigma} .
\end{align}
Substituting this into Eq.~\eqref{AiAnnAntiComt} we get
\begin{align}
    \{ \hat{z}_{i,\uparrow} \hat{z}_{i,\downarrow} , a_{i, \sigma}  \} 
     & = 2(2\hat{n}_{i, \bar{\sigma}} - \id )\left( a_{i, \sigma}  -   a_{i, \sigma} \right) \\
      & =0 .
\end{align}
For  $\{ \hat{z}_{i,\uparrow} \hat{z}_{i,\downarrow} , a^\dagger_{i, \sigma}  \}$, the argument is identical except now the relevant anti-commutator identity is $\{ \hat{n}_{i, \sigma} , a^\dagger_{i, \sigma}  \}=a^\dagger_{i, \sigma}$ which is obtained similarly to Eq.~\eqref{NumAnnAntiCom}.  

Next, to prove 
\begin{equation}
\{ \hat{z}_{i,\uparrow} \hat{z}_{i,\downarrow} ,B_{i,j} \}=0 ,
\end{equation}
we need only that $\hat{z}_{i,\uparrow} \hat{z}_{i,\downarrow}$ commutes with $a_{j, \sigma}$ and anticommutes with $a^\dagger_{i, \sigma}$.  When we expand out $B_{i,j}$, every term is of the form $a^\dagger_{i, \sigma} a_{j, \sigma}$ or $a_{i, \sigma} a^\dagger_{j, \sigma}$ for some $\sigma$ value.  Since $\hat{z}_{i,\uparrow} \hat{z}_{i,\downarrow}$ anticommutes with every term of $B_{i,j}$, it anticommutes with the full operator.

\section{Commutator bounds}

\subsection{Overview}
\label{AppCommutator}

Here we prove the following Trotter error bounds
\begin{theorem} \label{SquareTheorem}
    For the Hubbard model on an $L \times L$ square lattice with periodic boundary conditions, the second order trotter formaule obey
    \begin{align}
  || e^{i (H_h+H_I) t } - e^{i (t/2) H_h} e^{i tH_I}e^{i  (t/2) H_{\mathrm{h}}   }  ||  &   \leq W_{SO1} t^3  , \\ \nonumber
    || e^{i (H_h+H_I) t } - e^{i(t/2)  H_I} e^{i t H_h}e^{i (t/2) H_{I}  }  ||  &   \leq W_{SO2} t^3 ,
\end{align}
where 
\begin{align}
W_{SO1} & \leq  \left( \frac{u^2 \tau}{12} \right) || H_h || + \left( \frac{u \tau^2}{12} \right) L^2(\sqrt{5}+8) , \\ 
W_{SO2} & \leq \left( \frac{u \tau^2}{6} \right) L^2(\sqrt{5}+8) + \left( \frac{u^2 \tau}{24} \right) || H_h || .
\end{align}
Note that $|| H_h ||$ can be efficiently computed and we give some values in Table~\ref{HhopValues}.
\end{theorem}
Compared to the bounds numerically obtained by Kivlichan~\textit{et al.}~\cite{kivlichan2020improved} for the parameter choices $u/\tau = 4,8$, we find that our bounds are about 16 times better.  Note that, technically ours is a bound on a different Trotter sequence because we have shifted the chemical potential of the Hamiltonian. Unless $u \gg \tau$, the best bound is given by $W_{SO1}$.

\textit{Proof.-}  We make use of well known commutator bounds for second order Trotter~\cite{kivlichan2020improved,childs2019theory} so that
\begin{align} \nonumber
W_{SO1} & = \frac{||[ H_h ,H_I ],   H_I   ] ||}{12} + \frac{||  [ [ H_h , H_I ] , H_h   ] ||  }{24} ,
\end{align}
and
\begin{align} \nonumber
W_{SO2} & =  \frac{||  [[ H_I ,H_h ] ,  H_h   ] ||}{12} +  \frac{||  [ [ H_I , H_h ] , H_I   ]  ||}{24}  ,
\end{align}
and we can choose the minimum of the above two expressions since the gate-count is essentially the same for both orderings.

In App.~\ref{AppFirstLemma}, we present Lemma~\ref{FirstLemma} to obtain a bound on $||[ H_h ,H_I ],   H_I   ] ||$ .  In App.~\ref{AppSecondLemma}, we present Lemma~\ref{SecondLemma} and Corollary~\ref{MainCor} to obtain a bound on $||  [ [ H_I , H_h ] , H_I   ] ||$.  Combined these results gives bounds on $W_{SO1}$ and $W_{SO2}$

\setlength{\tabcolsep}{5pt}
\begin{table*}  
\begin{tabular}{lcccccccccccccccc} \toprule
$L$ & & 4 & 6 & 8 & 10 & 12 & 14 & 16 & 18 & 20 & 22 & 24 & 26 & 28 & 30 & 32 \\ \midrule
$\frac{1}{\tau}\|H_h  \| $ & & 24 & 56 & 100 & 160 & 230 & 320 & 410 & 520 & 650 & 780 & 930 & 1100 & 1300 & 1500 & 1700 \\
   $\frac{1}{\tau^3} \| [[H_h^r, H_h^g],H_h^r] \|  $ & & 0 & 110 & 190 & 300 & 440 & 630 & 810 & 100 & 130 & 160 & 180 & 220 & 250 & 290 & 330 \\ \bottomrule
\end{tabular}  \caption{Numerically computed values of normalised hopping Hamiltonian $\|H_h  \| / \tau$ and nested commutator of plaquette interaction Hamiltonians  for an $L \times L$ square, periodic Hubbard model.  This was found by using $\|H_h  \| = \tau || \tilde{R} ||_1$ where $\tilde{R}$ is the adjacency matrix for the $L \times L$ square, periodic lattice and similarly for the nested commutator.  We note that $\|H_h  \| / \tau \leq (3/2)L^2$ for all $L \geq 4$.} \label{HhopValues}
\end{table*}

\subsection{First lemma for Trotter error bounds}
\label{AppFirstLemma}

Here we prove the following Lemma
\begin{lem} \label{FirstLemma}
For any Hubbard-type Hamiltonian where $H_I=(u/4) \sum_{i} \hat{z}_{i,\uparrow} \hat{z}_{i,\downarrow}$ is the interacting Hamiltonian and $H_h$ is the hopping part, we have the following bound 
\begin{equation} \label{AppFirstBound}
  \|   [[ H_I ,H_h   ], H_I ] \|   \leq u^2  \|H_h  \| .
\end{equation}
\end{lem}
Furthermore, for the special case of the Hubbard model on an $L \times L$ square lattice with periodic boundary conditions, in Table~\ref{HhopValues} we provide precomputed values of 
$\|H_h  \| / \tau$ so that Eq.~\ref{AppFirstBound} can be easily evaluated. Here we present a proof of the Lemma in full generality. 

Throughout we use the notation
\begin{equation}  \label{eqn:HoppingApp}
H_h=\sum_{i \neq j} B_{i,j} ,
\end{equation}
where 
\begin{equation}
\label{eqn:HoppingApp2}
    B_{i,j}:= \sum_{\sigma}  R_{i,j}a^{\dagger}_{i,\sigma} a_{j,\sigma} .
\end{equation}  
Our first step is to compute $[ H_I ,H_h ]$. We see that $[\hat{z}_{k,\uparrow} \hat{z}_{k,\downarrow}, B_{i,j}]=0$ when $i,j,k$ are distinct and otherwise they anti-commute (see App.~\ref{AntiCommute}).  Therefore,
\begin{align}
\label{ABresult}
    [H_I, B_{i,j}] & = \frac{u}{4}  [\hat{z}_{i,\uparrow} \hat{z}_{i,\downarrow} + \hat{z}_{j,\uparrow} \hat{z}_{j,\downarrow}, B_{i,j}] \\ \nonumber
    & = \frac{u}{2} (\hat{z}_{i,\uparrow} \hat{z}_{i,\downarrow} + \hat{z}_{j,\uparrow} \hat{z}_{j,\downarrow})B_{i,j} .
\end{align}
Nesting the commutator again with $H_I$ and we get
\begin{align}
   [ [H_I, B_{i,j}] , H_I]  & = \frac{u}{2} [ (\hat{z}_{i,\uparrow} \hat{z}_{i,\downarrow} + \hat{z}_{j,\uparrow} \hat{z}_{j,\downarrow})B_{i,j} , H_I ]  \\ \nonumber & = \frac{u}{2} (\hat{z}_{i,\uparrow} \hat{z}_{i,\downarrow} + \hat{z}_{j,\uparrow} \hat{z}_{j,\downarrow})  [ B_{i,j} , H_I ] \\ \nonumber& = -\frac{u^2}{4} (\hat{z}_{i,\uparrow} \hat{z}_{i,\downarrow} + \hat{z}_{j,\uparrow} \hat{z}_{j,\downarrow})^2 B_{i,j} ,
\end{align}
where we have used that $[\hat{z}_{i,\uparrow} \hat{z}_{i,\downarrow},H_I]=0$ and we replaced $[ B_{i,j} , H_I ]$ using Eq.~\eqref{ABresult}.  Using that $(\hat{z}_{j,\uparrow} \hat{z}_{j,\downarrow})^2=\id$ and $[\hat{z}_{i,\uparrow} \hat{z}_{i,\downarrow},\hat{z}_{j,\uparrow} \hat{z}_{j,\downarrow}]=0$ we can evaluate the square to get
\begin{align} \label{partialABA}
   [ [H_I, B_{i,j}] , H_I ] & = -\frac{u^2}{2} (\id + \hat{z}_{i,\uparrow} \hat{z}_{i,\downarrow} \hat{z}_{j,\uparrow} \hat{z}_{j,\downarrow}) B_{i,j} .
\end{align}
We define the unitary 
\begin{equation}
 V_j = (\id +i \hat{z}_{j,\uparrow} \hat{z}_{j,\downarrow})/\sqrt{2}  ,
\end{equation}
and note that it satisfies
\begin{equation}
 (V_i \otimes V_j) B_{i,j}(V_i \otimes V_j)^\dagger   = - \hat{z}_{i,\uparrow} \hat{z}_{i,\downarrow} \hat{z}_{j,\uparrow} \hat{z}_{j,\downarrow} B_{i,j} .
\end{equation}
Furthermore, for $k \neq i,j$ the unitary $V_k$ commutes with $B_{i,j}$ so we can consider the unitary over all sites $\bar{V} := \prod_{j=1}^{n} V_j$, and we have
\begin{equation}
 \bar{V} B_{i,j} \bar{V}^\dagger   = - \hat{z}_{i,\uparrow} \hat{z}_{i,\downarrow} \hat{z}_{j,\uparrow} \hat{z}_{j,\downarrow} B_{i,j} .
\end{equation}
This enables us to rewrite Eq.~\eqref{partialABA} as follows
\begin{equation}
 [ [H_{I}, B_{i,j}] , H_{I} ]   = \frac{u^2}{2} (\bar{V} B_{i,j}\bar{V}^\dagger - B_{i,j} )  .
\end{equation}
Summing over all $i \neq j$, gives us
\begin{equation}
 [ [H_{I}, H_{h}] , H_{I}]   = \frac{u^2}{2} (\bar{V} H_{h} \bar{V}^\dagger - H_{h} )  .
\end{equation}
Taking the operator norm, using the triangle inequality and unitary invariance of the operator norm yields
\begin{equation}
\| [ [H_{I}, H_{h}] , H_{I}]     \| \leq u^2 || H_{h} || .
\end{equation}
Since $H_{h}$ is a free fermionic Hamiltonian the energy is known (recall Eq.~\eqref{Benergy}), so we have
\begin{equation}
\| [ [H_{I}, H_{h}] , H_{I}]  \| \leq  u^2 || R ||_1 ,
\end{equation}
where $R$ is the matrix of Hamiltonian coefficients from Eq.~\eqref{FullHop}.  This completes the proof of the Lemma~\ref{FirstLemma}.

\subsection{Second lemma for Trotter error bounds}
\label{AppSecondLemma}

Our second crucial lemma states that
\begin{lem} \label{SecondLemma}
For any Hubbard-type Hamiltonian where $H_I= (u/4) \sum_{i} \hat{z}_{i,\uparrow} \hat{z}_{i,\downarrow}$ is the interacting Hamiltonian and $H_h$ is the hopping part, we have the following bound 
\begin{equation}
 \|   [[ H_I , H_h   ], H_h ] \| \leq  \left( \frac{u}{2} \right) \sum_i \left( \|  [T_i,H_h  ]  \|  + 2 \| T_i \|^2 \right)
\end{equation}
where 
\begin{equation}
 T_{i} := \sum_{j : j \neq i} B_{i,j}.
\end{equation}
\end{lem}

To prove Lem.~\ref{SecondLemma}, 
we use the same abbreviated notation introduced in Eq.~\eqref{eqn:HoppingApp} and Eq.~\eqref{eqn:HoppingApp2} of Sec.~\ref{AppFirstLemma}. We begin by observing that
\begin{align}
       [[ H_I , H_h  ], H_h ] & = \sum_i \left( \frac{u}{4}\right)    [[ \hat{z}_{i,\uparrow} \hat{z}_{i,\downarrow}  , H_h  ], H_h ] . 
\end{align}
Taking the norm, using the triangle inequality, gives
\begin{align} \label{NormTaken}
   \| [[ H_I , H_h  ], H_h ] \| & \leq \left( \frac{u}{4}\right) \sum_i   \|  [[ \hat{z}_{i,\uparrow} \hat{z}_{i,\downarrow}  , H_h  ], H_h ]   \|  .
\end{align}
For a given vertex $i$, we have that
\begin{align}
    [ \hat{z}_{i,\uparrow} \hat{z}_{i,\downarrow} , H_h ] & = \sum_{j:j \neq i} [ \hat{z}_{i,\uparrow} \hat{z}_{i,\downarrow} , B_{i,j} ] , \\ \nonumber
    & = \sum_{j:j \neq i} 2 \hat{z}_{i,\uparrow} \hat{z}_{i,\downarrow}  B_{i,j}  .
\end{align} and we have used that for such indices the operators anticommute.  Furthermore, since $H_h=H_h^\dagger$ we have $B=\sum_{(i,j):i \neq j} B_{i,j}^\dagger$ and
\begin{align}
    [ \hat{z}_{i,\uparrow} \hat{z}_{i,\downarrow} , H_h ] &  = \sum_{j:j \neq i} 2 \hat{z}_{i,\uparrow} \hat{z}_{i,\downarrow}  B_{i,j}^\dagger .
\end{align}
Symmetrizing the above two results, we obtain
\begin{align}
    [ \hat{z}_{i,\uparrow} \hat{z}_{i,\downarrow} , H_h ] &  = \sum_{j:j \neq i}  \hat{z}_{i,\uparrow} \hat{z}_{i,\downarrow}  (B_{i,j}+B_{i,j}^\dagger)
\end{align}
Using the shorthand
\begin{equation}
  T_i := \frac{1}{2}  \sum_{j: j \neq i} (B_{i,j}+B_{i,j}^\dagger)  ,
\end{equation}
which represents all hopping terms that interact with site $i$, we have
\begin{align}
\label{ABvertex}
    [ \hat{z}_{i,\uparrow} \hat{z}_{i,\downarrow} , H_h ] & =  2 \hat{z}_{i,\uparrow} \hat{z}_{i,\downarrow}  T_i .
\end{align}
Nesting the commutator and using the identity $[PQ,R]=P[Q,R]+[P,R]Q$ with $P= \hat{z}_{i,\uparrow} \hat{z}_{i,\downarrow}  $, $Q=T_i$ and $R=H_h$, we have
\begin{align}
    [[ \hat{z}_{i,\uparrow} \hat{z}_{i,\downarrow} , H_h ] , H_h ] & =  2  \left[ \hat{z}_{i,\uparrow} \hat{z}_{i,\downarrow} T_i , H_h \right] \\ \nonumber
    & =  2 \hat{z}_{i,\uparrow} \hat{z}_{i,\downarrow}   \left[ T_i , H_h \right] + 2 \left[ \hat{z}_{i,\uparrow} \hat{z}_{i,\downarrow} , H_h \right] T_i  .
\end{align}
Using, Eq.~\eqref{ABvertex} we have
\begin{align}
    [[ \hat{z}_{i,\uparrow} \hat{z}_{i,\downarrow} , H_h ] , H_h ] &  =  2 \hat{z}_{i,\uparrow} \hat{z}_{i,\downarrow}  \left[ T_i , H_h \right]  + 4  \hat{z}_{i,\uparrow} \hat{z}_{i,\downarrow} T_i^2 . \nonumber 
\end{align}
Taking the operator norm, using the triangle inequality and $||\hat{z}_{i,\uparrow} \hat{z}_{i,\downarrow}||=1$ leads to
\begin{align}
   || [[ \hat{z}_{i,\uparrow} \hat{z}_{i,\downarrow} , H_h ] , H_h ] || &  \leq  2 ||  \left[ T_i , H_h \right] || + 4 ||  T_i^2 || \\
   &  =  2 ||  \left[ T_i , H_h \right] || + 4 || T_i ||^2 .
\end{align}
Summing over $i$ gives
\begin{align}
  \sum_i || [[ \hat{z}_{i,\uparrow} \hat{z}_{i,\downarrow} , H_h ] , H_h ] || \leq  \sum_{i} \left( 2 ||  \left[ T_i , H_h \right] || + 4 || T_i ||^2 \right).
\end{align}
and substituting into Eq.~\eqref{NormTaken} completes the proof of Lem.~\ref{SecondLemma}

The above lemma applies to all Hubbard type Hamiltonians.  In the special case of interest
\begin{corollary} \label{MainCor}
For the special case of the TDL Hubbard model on an $L \times L$ square lattice with periodic boundary conditions, we have
\begin{equation}
\| T_i \|=4 \tau ,
\end{equation}
and 
\begin{equation}
\| [T_i,H_h  ] \|= 4 \sqrt{5} \tau^2 .
\end{equation}
Therefore, we have
\begin{equation}
 \|   [[ H_I , H_h   ], H_h ] \| \leq u \tau^2  \left( 2 \sqrt{5} + 16 \right) L^2.
\end{equation}
\end{corollary}

For the square lattice, the operator $T_i$ is a hopping interaction corresponding to a star graph so
\begin{equation}
    R_{\mathrm{star}} = \tau \left( \begin{array}{ccccc}
       0  &  1 & 1 & 1 & 1 \\
       1  & 0 & 0 & 0 & 0 \\
       1  & 0 & 0 & 0 & 0 \\
       1  & 0 & 0 & 0 & 0 \\
       1  & 0 & 0 & 0 & 0 
    \end{array}\right) .
\end{equation}
We know from App.~\ref{AppFreeFermionic} that $||T_i||=||R_{\mathrm{star}}||_1$ and one easily finds that $||R_{\mathrm{star}}||_1=4 \tau$. 

To evaluate $\left[ T_i , B \right]$ we note that both operators are free fermionic, therefore the commutator is also a free fermionic operator with coefficients $R_{\mathrm{flower}} := [R_{\mathrm{star}}, R]$.  Since $R_{\mathrm{star}}$ acts locally on only 5 electrons, the resulting $||  \left[ T_i , H_h \right] ||=|| [R_{\mathrm{star}}, R] ||_1$ will be independent of the lattice size proved $L$ is large enough.  Computing this numerically for $L=5$, we get
\begin{equation}
    || [ T_i , H_h ] || =  4 \sqrt{5} \tau^2.
\end{equation}
Substituting this back into our nested commutator, we have
 \begin{align}
   || [[ \hat{z}_{i,\uparrow} \hat{z}_{i,\downarrow} , H_h ] , H_h ] || &  \leq  2 ||  \left[ T_i , B \right] || + 4 ||  T_i^2 || \\
   &  =  (8 \sqrt{5} + 4^3 ) \tau^2 ,
\end{align}
and so for an $L$ by $L$ lattice we have
 \begin{align}
       || [[ H_I , H_h ] , H_h ] || & \leq \frac{L^2 }{4} || [[ \hat{z}_{i,\uparrow} \hat{z}_{i,\downarrow} , H_h ] , H_h ] || \\ \nonumber
       & \leq u \tau^2 L^2 (2 \sqrt{5} + 16) .
\end{align}
which proves Corollary~\ref{MainCor}.

\section{PLAQ Trotter error}

The previous method requires a unitary that simultaneously diagonalizes all the hopping terms in $H_h$.  There are two previously proposed options for this: using Givens rotations or the Fast Fermionic Fourier Transform (FFFT).  Given rotations can be used on any size lattice, whereas the FFFT can only be performed on lattices of size $L=2^k$ with integer $k$.  Whereas, the Givens rotation method typically leads to much higher gate counts. 

We introduce plaquette Trotterization or simply PLAQ.  It is applicable to any lattice with even $L$, so is more versatile than the FFFT approach.  Furthermore, it provides gates counts that are comparable or superior to those obtained with FFFT, and far lower than using Givens rotations. We split $H_h$ into two components $H_h^p$ and $H_h^g$, so that
\begin{equation}
    H =  H_{I} + H_h^p + H_h^g ,
\end{equation}
as described in the main text. That is, we partition the edges into pink and gold  sets, such that every edge belongs to one set and the partitioning is regular.  This can be done straightforwardly on a periodic lattice if $L$ is even.  If $L$ is odd, then a similar results can be obtained by slightly breaking the regularity of the partitioning, but we ignore this case for brevity.  Already, all even $L$ is a much more fine grained than the SO-FFFT Trotterization scheme~\cite{kivlichan2020improved} that required $L=2^k$ with integer $k$.  Given this partitioning, we implement a single Trotter step as
\begin{equation}
    e^{-it H} \approx e^{-i(t / 2) H_I } e^{-i(t / 2) H_h^p }  e^{- i t H_h^g }   e^{-i(t / 2)  H_h^p } e^{-i (t / 2) H_I } .
\end{equation}

\begin{widetext}
Using well-known second order commutator bounds, we have that the error is bounded by $W_{\mathrm{PLAQ}} t^3$ where
\begin{equation*}
    W_{\mathrm{PLAQ}} =\frac{1}{12} \sum_{b=1,2} \left(||  \sum_{c>b, a>b} [[ H_b , H_c ], H_a ] ||  + \frac{1}{2} || \sum_{c>b} [[ H_b , H_c ], H_b] || \right)
\end{equation*}
with $H_1=H_I$, $H_2= H_h^p$ and $H_3= H_h^g$.  Performing the summations we get
\begin{align}
    W_{\mathrm{PLAQ}} & =\frac{1}{12}  \left(||  \sum_{c>1, a>1} [[ H_1 , H_c ], H_a ] ||  + \frac{1}{2} || \sum_{c>1} [[ H_1 , H_c ], H_1] ||  + \sum_{c>2, a>2} [[ H_2 , H_c ], H_a ] ||  + \frac{1}{2} || \sum_{c>2} [[ H_2 , H_c ], H_1] || \right) \\
    &  =\frac{1}{12}  \left(||  [[ H_1 , H_2 + H_3 ], H_2 + H_3 ] ||  + \frac{1}{2} ||  [[ H_1 , H_2 + H_3 ], H_1] ||  + || [[ H_2 , H_3 ], H_3 ] ||  + \frac{1}{2} ||  [[ H_2 , H_3 ], H_2] || \right) .
\end{align}
Noting that $H_2 + H_3 = H_h$ and $H_1=H_I$ we can simplify two terms to get
\begin{align}
    W_{\mathrm{PLAQ}}  &  = \frac{1}{12}  \left( u \tau^2 ||  [[ H_I , H_h ], H_h ] ||  + \frac{1}{2} u^2 \tau ||  [[ H_I , H_h ], H_I ] ||  +  || [[  H_h^p ,  H_h^g ], H_h^g ] ||  + \frac{1}{2} ||  [[ H_h^p ,H_h^g ],  H_h^p] || \right) .
\end{align}
\end{widetext}
Indeed, these first two terms of $W_{\mathrm{PLAQ}}$ exactly match the error contributions we calculated earlier, so
\begin{align}
    W_{\mathrm{PLAQ}}  &  = W_{SO2} +    W_{\mathrm{extra2}} ,
\end{align}
 and
\begin{align}
   W_{\mathrm{extra2}}  & =  \frac{1}{12} ||    [[ H_h^p , H_h^g ], H_h^g ]   || + \frac{1}{24} ||   [[ H_h^p , H_h^g ], H_h^p ]    ||  ,
\end{align}
represent the excess error from using plaquette Trotterization instead of split-operator Trotterization.  By symmetry between the pink and gold  sets, both commutators have the same norm and so 
\begin{align}
   W_{\mathrm{extra2}}  & =  \frac{3 }{24} ||    [[ H_h^p , H_h^g ], H_h^g ]   ||   ,
\end{align}
This is a nested commutator of free fermionic Hamiltonians.  Using $R^p$ and $R^g$ for the Hamiltonian coefficients, we have
\begin{align}
  W_{\mathrm{extra2}}  & =  \frac{3 }{24} ||    [[ H_h^p , H_h^g ], H_h^g ]   || \\
   & =  \frac{3 }{24} || [[ R^p , R^g], R^g] ||_1  ,
\end{align}
which can be efficiently computed. 

For $|| [[ R^p , R^g], R^g] ||_1 $ on the periodic, square $L \times L$ lattice, we present a precomputed set of values in Table~\ref{HhopValues} and note that across this range the values are upper bounded by $\frac{10}{3}L^2 \tau^3$.

\section{Realising plaquette Trotterization}
\label{App:realisePLAQ}

Here we give explicit circuits for realising the steps of the plaquette Trotterization and therefore count the number of non-Clifford gates required.

\subsection{Resource cost without ancilla}

First, we count gates without use of Hamming weight phasing. Given $r$ reps of 
\begin{equation}
    e^{i t H} \approx e^{i(t / 2) H_I } e^{i(t / 2) H_h^p }  e^{ i t H_h^g }   e^{i(t / 2)  H_h^p } e^{i (t / 2) H_I } ,
\end{equation}
we can merge the first and last terms
\begin{equation}
    e^{i  t H} \approx  e^{i(t / 2) H_I } [ e^{i(t / 2) H_h^p }  e^{ i t H_h^g }   e^{i(t / 2)  H_h^p } e^{i t H_I }]^r e^{-i(t / 2) H_I } ,
\end{equation}
so for large $r$ only the terms in the square brackets are non-negligible.  We will describe costs using a vector $\vec{N}=\{ N_{\mathrm{TOF}}, N_T, N_R \}$ that represent the number of Toffoli gates, $T$ gates and arbitrary single qubit $Z$ rotations. Note that these costs are per Trotter step. There are various conversions between these costs that we discuss later.

For each step, we make 1 implementation of the interaction step $e^{-i t H_I }$ and 3 implementations of tile steps $ e^{-i(t / 2) H_h^p }$ or $e^{- i t H_h^g }$ (the cost of which is equal), so 
\begin{equation}
    \vec{N} = \vec{N}_{\mathrm{int}} + 3   \vec{N}_{\mathrm{tile}}. 
\end{equation}
There are $L^2$ interactions, but because of our choice of chemical potential each interaction corresponds to 1 arbitrary rotation, so it is possible to realise this with cost $\vec{N}_{\mathrm{int}} = \{0,0, L^2 \}$.  Without our choice of chemical potential, we would have counted three times as many gates. 

Next, we consider implementation of $ e^{-i(t / 2) H_h^p }$ or $e^{- i t H_h^g }$.  The lattice contains $L^2/4$ plaquettes of each colour.  Accounting for spin that makes $L^2/2$ plaquettes inside one spin sector, so we can write $\vec{N}_{\mathrm{tile}}= (L^2/2) \vec{N}_{\mathrm{plaq}}$. Here $\vec{N}_{\mathrm{plaq}}$ represents the cost of realising unitaries of the form $\exp(i \tau K)$ where $K$ is a plaquette operator
\begin{equation}
    K = \sum_{i,j} [R_{\mathrm{plaq}}]_{i,j} a_i^\dagger a_j ,
\end{equation}
where we drop the spin index for brevity and the non-zero sub-block of $R_{\mathrm{plaq}}$ is
\begin{equation}
    R_{\mathrm{plaq}} = \left(  \begin{array}{cccc}
       0  & 1 & 0 & 1 \\
       1 & 0 & 1 & 0 \\ 
       0  & 1 & 0 & 1 \\
       1 & 0 & 1 & 0 \\ 
    \end{array}
    \right) .
\end{equation}
To see this, note that a square is a small periodic loop of size 4.  Labeling the fermions going around the loop, we have that $i$ and $j$ interact if those numbers differ by 1 (modulo 4), which leads to $R$ above.  We can diagonalize $R_{\mathrm{plaq}}$ to find the Givens rotation that diagonalizes $\exp(i \tau  K)$.  In general, the cost would be 3 Givens rotations, 4 arbitrary rotations, and the inverse Givens rotations. However, when we diagonalize $R_{\mathrm{plaq}}$ we find that it has only two non-trivial eigenvalues, which will dramatically reduce the gate count.  That is,
\begin{equation}
    K = 2 b^\dagger b - 2 c^\dagger c ,
\end{equation}
where
\begin{align}
   b & = (a_1 + a_2 + a_3 + a_4 )/2 , \\ \nonumber
   c & = (a_1 - a_2 + a_3 - a_4 )/2 ,
\end{align}
Letting $F_{i,j}$ be a fermionic operator such that
\begin{align}
   F_{i,j} a_i F_{i,j}^\dagger & = (a_i+a_j)/\sqrt{2} , \\ \nonumber
   F_{i,j} a_j F_{i,j}^\dagger & = (a_i-a_j)/\sqrt{2} . 
\end{align}
Then using $V= F_{3,1} F_{2,4} F_{2,3} $ we get
\begin{align}
     V a_2 V^\dagger & = b , \\ \nonumber
     V a_3 V^\dagger & = c  .
\end{align}
Note that the qubit implementation of $F_{i,j}$ is non-local when $i$ and $j$ are non-adjacent with respect to the Jordon-Wigner string ordering.  However, we can always use fermionic swap $f_{\mathrm{SWAP}}$ operations to make modes adjacent, and since fermionic swap can be realised using Clifford operations we neglect this cost.  This is valid for the resource costing model used in this paper.  However, additional care is required when costing in other settings.

Next, we can realise the plaquette time evolution via
\begin{align}
\exp(i \tau  K) & = V  e^{i 2  \tau a_2^\dagger a_2} e^{-i 2  \tau a_3^\dagger a_3}     V^\dagger \\ \nonumber
& = F_{3,1} F_{2,4} F_{2,3}  e^{i 2  \tau a_2^\dagger a_2} e^{-i 2  \tau a_3^\dagger a_3 } F_{2,3}^\dagger F_{2,4}^\dagger    F_{3,1}^\dagger  
\end{align}
However, we can compile further.  Assuming a Jordon-Wigner encoding, we have that $F_{2,3}$ is realised by
\begin{equation}
    F_{2,3} = \left( \begin{array}{cccc}
       1  & 0 & 0 & 0  \\
       0  & 1/\sqrt{2} & 1/\sqrt{2} & 0 \\
       0  & 1/\sqrt{2} & -1/\sqrt{2} & 0 \\
       0  & 0 & 0 & -1  \\
    \end{array}\right) ,
\end{equation}
and 
\begin{equation}
 e^{i 2 t \tau a_2^\dagger a_2} e^{-i 2 t \tau a_3^\dagger a_3 }     = \left( \begin{array}{cccc}
       1  & 0 & 0 & 0  \\
       0  & e^{ 2 i  \tau} & 0 & 0 \\
       0  & 0 & e^{- 2  t \tau } & 0 \\
       0  & 0 & 0 & 1  \\
    \end{array}\right),
\end{equation}
we find that  
\begin{align} \nonumber
    F_{2,3}  e^{i 2 t \tau a_2^\dagger a_2} e^{-i 2 t \tau a_3^\dagger a_3 } F_{2,3} & = \left( \begin{array}{cccc}
    1  & 0 & 0 & 0  \\
    0  & \cos(2\tau) & i \sin(2\tau) & 0  \\
    0   & i \sin(2\tau) & \cos(2\tau)  & 0  \\
   0  & 0 & 0 & 1  \\ 
    \end{array}
    \right) \\ 
   & = e^{i \tau X \otimes X} e^{i \tau Y \otimes Y} . 
\end{align}
Using a Clifford we can rotate $X \otimes X$ and $Y \otimes Y$ to $Z \otimes \id$ and $\id \otimes Z$.  Therefore, each plaquette unitary requires 2 arbitrary $Z$ rotations (of equal angle) and 4 $F$ gates.   Each $F$ gate can be implemented using 2 $T$ gates (see Fig 8 of Ref.~\cite{kivlichan2020improved} using $k=0$, and noting that we use the subscripts of $F$ to indicate the fermion labels). Therefore, we have $\vec{N}_{\mathrm{plaq}}=\{ 0, 8, 2 \}$, which entails $\vec{N}_{\mathrm{tile}}= \{ 0, 4 L^2,  L^2 \}$.

The total cost is therefore,
\begin{align} \nonumber
    \vec{N} & = \vec{N}_{\mathrm{int}} + 3   \vec{N}_{\mathrm{tile}} \\ \nonumber
    & = \{  0, 0, L^2 \} + 3\{ 0, 4 L^2,  L^2 \}\\ \nonumber
    & = \{ 0, 12 L^2, 4 L^2 \}.
\end{align}
Recall that single qubit rotations require compilation into $T$-gates.  Typically, the optimal choice is roughly 40 $T$-gates per arbitrary rotation, which would lead to a cost \begin{equation}
    \vec{N} \approx  \{ 0, 172 L^2, 0 \} ,
\end{equation}
where the cost of synthesis is a significant factor.  However, within each factor of the Trotterization, all the arbitrary rotations are diagonalized to $Z_j$ rotations by the same angle.  Therefore, given access to ancillas we can use Hamming weight phasing to significantly reduce the total number of non-Cliffords.

\subsection{Using ancilla and Hamming weight phasing}
\label{App:HWP}

Hamming weight phasing is a gadget that uses ancilla and Toffoli gates to reduce the number of rotations needed.  
\begin{theorem}
Using Hamming weight phasing, it is possible to implement $m$ phase gates $\prod_{j=1}^m \exp(i \theta Z_{j})$ using 
$k=Floor[Log[2, m] + 1]$ arbitrary rotations, $\alpha$ Toffoli gates and $\alpha$ clean ancilla, where
\begin{equation}
\label{HammingWeight}
    \alpha = m - w(m) \leq m-1 ,
\end{equation}
where $w(m)$ is the number of 1s in the binary expansion of $m$.
\end{theorem}  
This is a slight refinement of the idea introduce by Gidney~\cite{gidney2018halving} and expanded up to in Ref.~\cite{kivlichan2020improved}.  Previously, it has been noted that Hamming weight phasing requires $\alpha \leq m-1$ and that this will be loose when $m \neq 2^q$ for some integer $q$. 

Our proof of the above claim combines two previously established facts. It has been observed that for quantum circuits, the Toffoli and ancilla cost of computing (and then uncomputing) some Boolean function is captured by the classical notion of multiplicative complexity~\cite{meuli2019role}.  Furthermore, it has been shown~\cite{boyar2005exact} that multiplicative complexity of Hamming weight computations is precisely captured by Eq.~\eqref{HammingWeight}.  Note that in Ref.~\cite{boyar2005exact} they use the notation $ H^{\mathbb{N}}(m)$ where we use $w(m)$.

The PLAQ Trotter step contains layers of $L^2$ or $L^2/2$ identical rotations.  If $m$ is a factor of $L^2/2$, we can break up the task in batches that each use Hamming weight phasing to transform the cost model as follows:  
\begin{align}
    \vec{N} & = \{0, 12L^2 , 4L^2  \}  \\
    & \rightarrow \{4 L^2 \alpha / m  , 12 L^2 , 4 L^2 \mathrm{Floor}[ \log_2(m) + 1] /m \}_t \nonumber
\end{align}
where $\alpha = m - w(m)$ and we assume access to $\alpha$ logical ancilla.  

If we convert each Toffoli gate into 4 $T$ gates,  we get a total cost of $\vec{N}=\{ 0, N_T, N_R \}$ where
\begin{align}
    N_T & = \left( \frac{16 \alpha}{m} + 6 \right)L^2  ,\\  \nonumber
    N_R & = \frac{4}{m} \mathrm{Floor}[ \log_2(m)+1] L^2  , \nonumber
\end{align}
In the worst case $\alpha=m-1$ and $\mathrm{Floor}[ \log_2(m)+1]=\log_2(m)+1$ for which
\begin{align}
    N_T & = \left( \frac{16 (m-1)}{m} + 12 \right)L^2  ,\\  \nonumber
    N_R & = \frac{4}{m} [ \log_2(m)+1] L^2  , \nonumber
\end{align}
Since the $N_R$ gates require further synthesis, HWP significantly improves the $T$-count.  However, this must be balanced against cost of the extra logical ancilla.

\section{Phase estimation overview}
\label{AppPhaseEstimation}

The phase estimation protocol has three different sources of additive error in the estimated energy.  Given a $2^{\mathrm{th}}$ order Trotter-Suzuki formula for unitary $U$, one will have
\begin{equation}
    || e^{i H t} - U_{TS} || \leq W t^3 ,
\end{equation}
where $W$ is a constant depending on the Hamiltonian $H$ and the exact type of Trotterization used.  We give explicit bounds on $W$ in Eqs.~\eqref{Wbounds1} and~\eqref{Wbounds2} of the main text (see also Thm.~\ref{SquareTheorem} of App.~\ref{AppCommutator}).  When used for phase estimation this leads~\cite{kivlichan2020improved} to an error in the energy of size
\begin{equation}
    \Delta_{TS} \leq \frac{1}{t}  || e^{i H t} - U_{TS} ||\leq W t^2 ,
\end{equation}
provided that $t$ is sufficiently small (which is verified in all our calculations).  The phase estimation protocol used by Kivlichan~\cite{kivlichan2020improved} contributes an error
\begin{equation} 
    \Delta_{PE} = \frac{0.76 \pi}{N_{PE} t} .
\end{equation}
where $N_{PE}$ counts the total applications of unitary $U_{TS}$.  The third source of error comes from the synthesis of single qubit $Z$ rotations into Clifford+$T$ gates, and we have that if a single Trotter step contains $N_R$ such rotations, we can achieve $\Delta_{HT}$ error if each rotation has a $T$-count of
\begin{equation}
    N_{HT} = 1.15 \log_2 ( N_R /( \Delta_{HT} t ) ) + 9.2.
\end{equation}
Due to the logarithmic dependence, optimal solutions typically set $\Delta_{HT} \ll \Delta_{TD}+\Delta_{PE}$.  For this reason, we first ask how best to choose $t$ to minimise  $\delta := \Delta_{TD}+\Delta_{PE}$.  The minima of $\delta$ can be found by differentiation w.r.t to $t$, which leads to 
\begin{equation}
    t = \left( \frac{0.76 \pi}{2 W N_{PE}} \right)^{1/3} .
\end{equation}
Substituting back into $\delta$, we get
\begin{equation}
    \delta = \Delta_{TS}+\Delta_{PE} = 3 \frac{W^{1/3}}{N_{PE}^{2/3}}\left( \frac{ 0.76 \pi}{2 }  \right)^{2/3}.
\end{equation}
and it is worth noting that $\Delta_{TS}= \frac{1}{3}\delta$ and $\Delta_{PE}= \frac{2}{3}\delta$.

Therefore, $N_{PE}$ must be set to an integer larger than 
\begin{equation}
    N_{PE} = \left( \frac{3^{3/2} 0.76 \pi }{ 2 } \right)  \frac{W^{1/2}}{\delta^{3/2}} \approx 6.203 \frac{W^{1/2}}{\delta^{3/2}} .
\end{equation}
\begin{widetext}
Let us consider the case when Toffoli gates are performed using $T$-gates. If each Trotter step uses $N_T$ direct $T$-gates and $N_R$ arbitrary $Z$-axis rotations (to be synthesized), then the total $T$ gate cost will be
\begin{align}
    N_{\mathrm{tot}} &= N_{PE}( N_R N_{HT} + N_T) \\ \nonumber
    & \approx 6.203 \frac{W^{1/2}}{\delta^{3/2}} \left( N_R\left(1.15 \log_2 \left( \frac{N_R}{\Delta_{HT} t} \right) + 9.2 \right) + N_T \right) .
\end{align}
Using $\Delta_{TS}= \frac{1}{3}\delta$ and $\Delta_{TS}= W t^2$, we know $t^2 = \delta / (3W)$.  
Substituting this in gives
\begin{align}
    N_{TOTAL} &= N_{PE}( N_R N_{HT} + N_T) \\ \nonumber
    & \approx 6.203 \frac{W^{1/2}}{\delta^{3/2}} \left( N_R\left(1.15 \log_2 \left( \frac{N_R (3W)^{1/2}}{ \Delta_{HT} \sqrt{\delta}} \right) + 9.2 \right) + N_T \right) .
\end{align}
Given a target error $\epsilon$ we require that $\Delta_{HT} + \delta \leq \epsilon$ and it remains to find the optimal split.  For instance, we may write $\delta= (1-x)\epsilon$ and $\Delta_{HT}=x \epsilon$, so we get
\begin{align}
    N_{TOTAL} &= N_{PE}( N_R N_{HT} + N_T) \\ \nonumber
    & \approx 6.203 \frac{W^{1/2}}{(1-x)^{3/2}\epsilon^{3/2}} \left( N_R\left(1.15 \log_2 \left( \frac{N_R (3W)^{1/2}}{ x  \sqrt{1-x} \epsilon^{3/2} } \right) + 9.2 \right) + N_T \right) .
\end{align}
Given a particular circuit implementation, the only free parameter left is $x$.  Numerically, we find that $x \sim 0.01$ is a good choice for the Hubbard model.  Above $N_{TOTAL}$ gives the total number of $T$-gates.  A similar calculation can be performed to instead count the total number of Toffoli gates through using catalysis. 
\end{widetext}

\end{document}